# Resource Sharing and Pipelining in Coarse-Grained Reconfigurable Architecture for Domain-Specific Optimization

Yoonjin Kim[†], Mary Kiemb[‡], Chulsoo Park[†], Jinyong Jung[†], Kiyoung Choi[†‡]

Design Automation Laboratory, School of EE/CS
Seoul National University, Seoul, South Korea
[†]{ykim, jupiter, jyjung}@poppy.snu.ac.kr, [‡]kiemb@marykiemb.net, [†‡]kchoi@azalea.snu.ac.kr

**Abstract**

*Coarse-grained reconfigurable architectures aim to achieve both goals of high performance and flexibility. However, existing reconfigurable array architectures require many resources without considering the specific application domain. Functional resources that take long latency and/or large area can be pipelined and/or shared among the processing elements. Therefore the hardware cost and the delay can be effectively reduced without any performance degradation for some application domains. We suggest such reconfigurable array architecture template and design space exploration flow for domain-specific optimization. Experimental results show that our approach is much more efficient both in performance and area compared to existing reconfigurable architectures.*

## 1 Introduction

As the market pressure of embedded systems compels the designer to meet tighter constraints on cost, performance and power, the application specific optimization of a system becomes inevitable. On the other hand, the flexibility of a system is also important to accommodate rapidly changing consumer needs. To compromise these incompatible demands, domain-specific design is focused on as a suitable solution for recent embedded systems. Coarse-grained reconfigurable architecture is the very domain-specific design in that it can boost the performance by adopting specific hardware engines but it can be reconfigured as well to adapt the different characteristics of each application.

In this reason, many delicate coarse-grained reconfigurable designs are proposed [1]. Most of them comprise of a fixed set of specialized processing elements (PEs) and interconnection fabrics between them and the run-time control of the operation of each PE and the interconnection provides the reconfigurability. However, such fixed architectures have limitations in optimizing the cost and performance for various domains of application. Some researchers suggest a reconfigurable architecture in the form of a template to find an optimal design for a specific application domain [4][5].

Most design space exploration techniques previously suggested are limited to the configuration of the internal structure of a PE and the interconnection scheme. Such configuration techniques are in general good at obtaining high performance but require high hardware cost. This is mainly because even a primitive PE design should be equipped with basic functional resources to gain reasonable performance. Moreover, adding a small functional block to a primitive PE design increases the total cost of the aggregate architecture increases a lot. To alleviate this problem, some templates permit heterogeneous PEs[5][6]. However, such heterogeneity bears strict constraints in the mapping of applications to PEs, resulting in negative effect on the performance.

In this paper, we suggest a reconfigurable architecture template which has two key features. One is hardware cost reduction by sharing critical functional resources that occupy large area in the PEs and the other is critical path reduction by pipelining the critical resources. To implant these features into our template, we assume that the template is based not on SIMD like execution [2] but on loop pipelining execution [7][8]. Although SIMD like execution is efficient for data parallel computation in that it saves configuration and data storage by regular execution, it is inflexible in that each PE cannot operate independently. In contrast, loop pipelining requires more storage to control each pipeline stage but it has more flexibility in selecting the operation of a PE. In addition, it can enhance the performance because it reduces the synchronization overhead to perform identical operations simultaneously.

This paper is organized as follows. After mentioning the related work in Section 2, we describe our reconfigurable architecture template in Section 3. In Section 4, we propose a design flow to optimize the reconfigurable architecture to a specific application domain. We also describe the details of design space exploration for optimal resource sharing. We show the experimental results in Section 5 and conclude with future work in the last section.

## 2 Related Work

Many coarse-grained reconfigurable architectures are suggested as summarized in [1]. Among them, two dimensional mesh architectures have the advantage of rich communication resources for parallelism. Morphosys [2] consists of 8 x 8 array of Reconfigurable Cell coupled with Tiny_RISC processor. The array performs 16-bits operations including multiplication in SIMD style. It shows good performance for regular code segments in computation intensive domains but requires high hardware cost. XPP configurable system-on-chip architecture [3] is another example. XPP has 4 x 4 or 8 x 8 reconfigurable array and LEON processor with AMBA bus architecture. A processing element of XPP is composed of an ALU and some registers. Since the processing elements do not include heavy resources, the total hardware cost is not high but the range of applicable domains is restricted.

For more aggressive domain-specific optimization, template based architectures are suggested. In ADRES template [4], an XML-based architecture description language is used to define the overall topology, supported operation set, resource allocation, timing, and even internal organization of each processing element. KressArray [5] also defines the exploration properties such as array size, interconnections, and functionality of certain processing elements. However, both templates do not support common resources shared among processing elements, thus some critical functional resources may have low utilization while occupying large area.

## 3 Architecture Template

### 3.1 Resource sharing

We illustrate resource sharing between processing elements with matrix multiplication. We assume a general mesh-based coarse-grained reconfigurable array of PEs, where a PE is a basic reconfigurable element composed of an ALU, an array multiplier, etc. and the configuration is controlled by configuration cache. Each row of the array shares read/write-buses. Figure 1 shows the case of 4x4 array with two read buses and one write-bus.

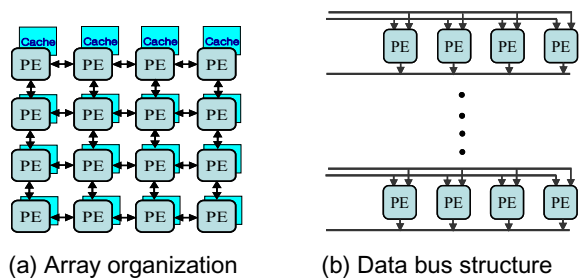

(a) Array organization    (b) Data bus structure

**Figure 1.  4x4 reconfigurable Array.**

Consider two square matrices X and Y of order N. The output matrix Z is represented by:

$$Z(i, j) = C \times \sum_{k=0}^{N-1} \{X(i,k) \times Y(k, j)\} \quad (1)$$

where i, j = 0,1,…,N and C is a constant specified in the configuration cache.

|       | 1  | 2  | 3  | 4  | 5  | 6  | 7  | 8  |
|-------|----|----|----|----|----|----|----|----|
| col#1 | Ld | *  | +  | +  | *  | St | Ld | *  |
| col#2 |    | Ld | *  | +  | +  | *  | St | Ld |
| col#3 |    |    | Ld | *  | +  | +  | *  | St |
| col#4 |    |    |    | Ld | *  | +  | +  | *  |

Ld: load operation, St: store operation

**Figure 2. The loop pipelining of a matrix multiplication of order 4.**

Assuming that equation (1) with N=4 is executed on the array shown Figure 1, the loop pipelining [7] schedules the operations as shown in Figure 2. The first row of Figure 2 represents the schedule time in cycle from the start of the loop and the first column represents the column number of the array. At the first cycle, all the PEs in the first column are loaded with the operands then perform multiplication at the second cycle. In the next two cycles, the PEs in the first column perform addition to obtain the sum of products, while the PEs in the next columns perform multiplication. Then multiplication is performed in the first column at the 5$^{th}$ cycle while the first multiplication is performed in the fourth column.

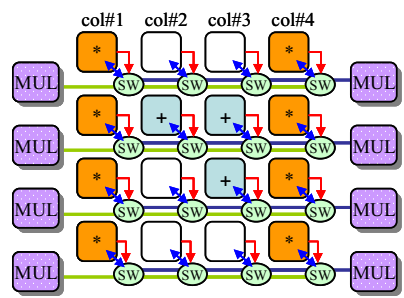

**Figure 3. 8 multipliers sharing among 16 PEs.**

Because loop pipelining distributes the same operations over several cycles, there is no need for all PEs to have the same functional resources at the same time. This allows the PEs in the same column or in the same row to share area-critical resources. For the 4x4 matrix multiplication example, since multipliers take much more area and delay compared to other resources, we classify them as critical resources and other resources as primitive resources. The reconfigurable structure in our template can be one dimensional array, two-dimensional mesh, or other rectangular structure, where each PE has its own Bus switch to control the resource sharing.

Figure 4 depicts the detailed connections about multiplier sharing. The two n-bit operands of a PE are connected to the Bus switch. The dynamic mapping of a mul-tiplier to a PE is determined in compile time and the information is annotated to the configuration instructions. In run-time, the mapping control signal from the configuration cache is fed to the Bus switch and the Bus switch decides where to route the operands. After the multiplication, the 2n-bit output is transferred from the multiplier to the original issuing PE via the Bus switch.

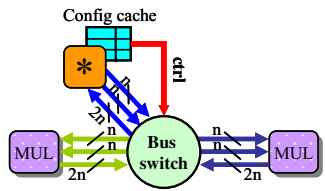

**Figure 4. The connections between a PE and shared multipliers.**

### 3.2 Resource pipelining

If there is a critical functional resource which has very long latency in a PE, the functional resource can be pipelined to curtail the critical path. Resource pipelining has clear advantage in loop pipelining execution because heterogeneous functional units can run at the same time. Figure 5 shows this. In a general PE design, the latency is fixed by the critical path but in a pipelined PE design, the critical path is curtailed by the register insertion, so the latency can vary depending on the operation. Since heterogeneous operations are executed simultaneously in loop pipelining, some short latency operations can terminate early and the total latency of the system may be reduced.

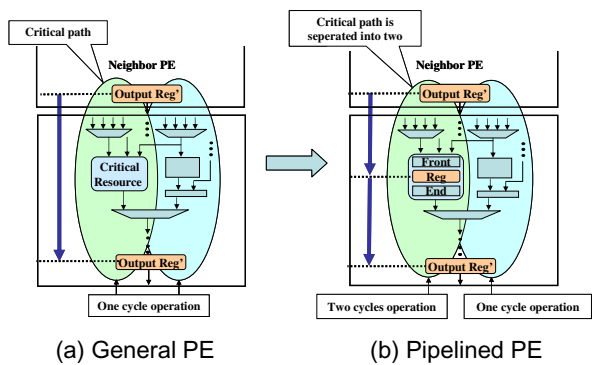

**Figure 5. The critical path comparison between a general PE and a pipelined PE.**

If a critical functional resource has both large area and long latency, the resource sharing and resource pipelining can be applied at the same time in such a way that the shared resource executes operation in several pipeline stages. If these two techniques are merged, the conditions for resource sharing are relaxed and so the critical resources are utilized more efficiently. Figure 6 shows this situation. It is the scheduling for equation (1) when the multiplier is pipelined into two stages. It needs only 4 multipliers to perform the execution without any stall whereas the scheduling of Figure 2 needs 8 multipliers. This is because two PEs sharing one pipelined multiplier can perform two multiplications at the same time using different pipeline stages.

|  | 1 | 2 | 3 | 4 | 5 | 6 | 7 | 8 | 9 | 10 |
|---|---|---|---|---|---|---|---|---|---|---|
| col#1 | Ld | 1* | 2* | + | + | 1* | 2* | St | Ld | 1* |
| col#2 |  | Ld | 1* | 2* | + | + | 1* | 2* | St | Ld |
| col#3 |  |  | Ld | 1* | 2* | + | + | 1* | 2* | St |
| col#4 |  |  |  | Ld | 1* | 2* | + | + | 1* | 2* |

1*/2* : first/second stage of pipelined multiplication

**Figure 6. The loop pipelining of a matrix multiplication when the multiplier is pipelined.**

## 4 Design Space Exploration

Since our Resource Sharing and Pipelining (RSP) technique assumes any rectangular pipelining structure, our RSP design flow for domain-specific optimization is applicable to a generic array architecture template including many existing coarse-grained reconfigurable arrays. The RSP technique can be integrated into the refinement stage of the base design space exploration flow. The steps in the upper half of Figure 7 represent the generic design space exploration flow and those in the lower half represent our RSP flow. The upper half can be slightly different depending on the specific base architecture.

To extract critical loops which are to be executed on the reconfigurable architecture, profiling is performed over a set of applications in the target domain. When the critical loops are selected, the design space is explored to obtain an optimal base architecture. After the base architecture is determined, the selected critical loops are mapped on the configuration contexts. With the base architecture and the initial configuration contexts, RSP parameters are determined through the RSP design space exploration step.

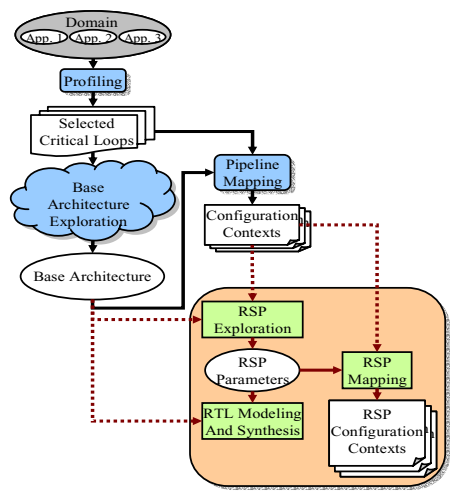

**Figure 7. Design space exploration flow for RSP architecture template.**

Our RSP template has the following principal parameters for design space exploration:
- The types of shared functional resources
- The types of pipelined resources
- The number of pipeline stages of the pipelined resources
- The number of rows of the shared resources
- The number of columns of the shared resources

For an easy mapping of operations, we need to form a regular structure of shared resources. So the shared resources are placed in line with the rows and/or columns of the reconfigurable structure. Thereby, the structure of the shared resources is represented by the number of rows and/or columns.

The RSP design space exploration is based on the estimation of the hardware cost and performance with the RSP parameters. If the hardware cost is too high or the performance is too low, such RSP design is rejected. Among the designs that satisfy the condition, only Pareto points are evaluated and then an optimal solution is selected. Although accurate hardware cost of entire architecture is evaluated after RSP exploration, we can estimate the hardware cost of an RSP design with pre-synthesized architecture components during RSP exploration. This is as follows :

$$HW_{cost} = \{n \times m \times (Sh\_PE_{area} + Reg_{area} + SW_{area}) + Sh\_Res_{area} \times (n \times shr + m \times shc)\} < n \times m \times PE_{area} \quad (2)$$

where

$n, m$ : number of row, column

$PE_{area}$ : area of a PE with shared resource

$Sh\_PE_{area}$ : area of a PE without shared resource

$Reg_{area}$ : area of registers in a PE for pipelined operation

$SW_{area}$ : area of a bus switch

$Sh\_Res_{area}$ : area of a shared resource

$shr/shc$ : number of rows/columns of the shared resources

Once the final RSP configuration context is obtained, we can evaluated the exact performance since the configuration context has the information of the exact number of cycles required to execute the given kernel loop. However, the mapping and evaluation of all the candidate RSP designs are time-consuming. Therefore, in the RSP exploration stage, we use the upper bound for the performance estimation. The upper bound is computed with the RSP parameters and the initial configuration contexts. In the case of RS, the number of operations that are to be executed on the critical resources is counted and compared with the number of shared resources in every cycle. If the number of critical operations in a cycle is larger than the number of shared resources, it means that the shared resources are lacking in the cycle. Therefore, some stall cycles should be inserted to avoid the resource conflict – we call it RS stall. In addition, in the case of RP, the operations that are to be executed on the pipelined resources take multiple cycles so the operations dependent on pipelined resources should be stalled together – we call it RP stall. So we approximate the entire number of stall cycles by RS stall and RP stall cycles. In reality, more cycles may stall depending on the characteristics of the RSP structure, thus this approximation is an upper bound of the performance.

The initial RSP configuration contexts are rearranged according to the RS and RP. We have two rules for the rearrangement. First, in the case of RS, shared resources are assigned to PEs in the order of loop iteration. Therefore, if it lacks shared resources, the operations in later loop iterations are moved to the next cycle. Second, in the case of RP, since the operations on pipelined resources take multiple cycles, other operations dependent on the output of pipelined resources have to be stalled together. Furthermore, in the case of consecutive pipelined operations, overlapped cycles between the operations should be removed. In the case of RSP, the two rules are applied to the initial configuration contexts.

## 5 Experimental Results

### 5.1 Architecture specification

We have determined the internal PE design as well as the array structure by analyzing the kernels and implemented the base architecture with VHDL. The base architecture used in this experiment is similar to Morphosys, which has a two dimensional 8 x 8 mesh of PEs. Each of the PEs contains one ALU and one array multiplier. Our base architecture is different from Morphosys in that data memory have multiple read/write data buses and a configuration cache is allocated to each PE. This is because our template assumes loop pipelining style execution while Morphosys assumes SIMD style execution. In addition, the bit-width of the data bus is extended to 16 and some interconnections between PEs are added to reduce data arrangement cycles. We evaluated area and delay cost of each component of a PE by RTL synthesis and the result is shown in Table 1. The array multiplier is a critical resource in both area and delay and thus we extract the multiplier from the PE design and arrange it to be the shared and pipelined resource. To make a two stage pipelined version of the multiplier, we inserted a pipeline register into the multiplier.

**Table 1. Synthesis result of a PE**

| Component | Area | | Critical path delay | |
|---|---|---|---|---|
| | No. of slices | Ratio | Time(ns) | Ratio |
| PE | 910 | 100 | 25.6 | 100 |
| Multiplexer | 58 | 6.37 | 1.3 | 12.89 |
| ALU | 253 | 27.80 | 11.5 | 44.92 |
| Array multiplier | 416 | **45.71** | 19.7 | **76.95** |
| Shift logic | 156 | 17.14 | 2.5 | 17.58 |

**bold number** : the largest ratio among the components

## 5.2 Architecture evaluation

For quantitative evaluation of RTL model from RSP exploration, we have used Synplify Pro[TM][9] as the RTL synthesis tool and Xilinx Virtex II 8M gates FPGA[10] as the target hardware. To demonstrate the effectiveness of our approach, we have compared three cases: base architecture, RS architecture, and RSP architecture. Figure 8 shows four cases of resource sharing:

1. one multiplier shared by 8 PEs in each row,
2. two multipliers shared by 8 PEs in each row,
3. two multipliers shared by 8 PEs in each row and one multiplier shared by 8 PEs in each column,
4. two multipliers shared by 8 PEs in each row and two multipliers shared by 8 PEs in each column.

The RTL synthesis results of these designs are shown in Table 2. Compared to the base architecture, we have reduced the area and delay by up to 42.8% and 34.69%, respectively.

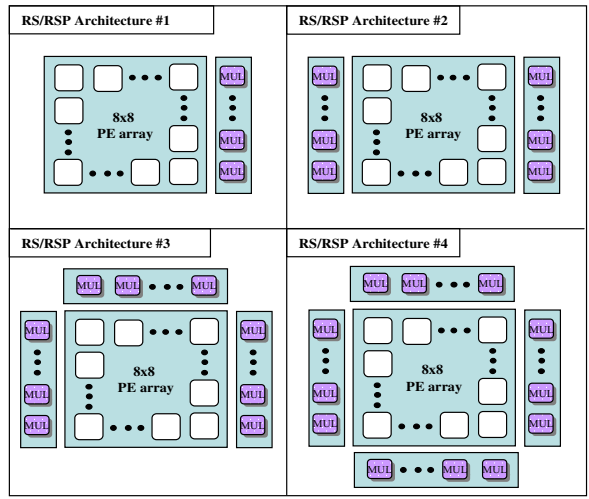

**Figure 8. Four designs of RS/RSP architectures.**

**Table 2. Synthesis result of various architectures**

| Arch' | Area (No. of slices) | | | | Critical path delay(ns) | | | |
|---|---|---|---|---|---|---|---|---|
| | PE | SW | Array | R(%) | PE | SW | Array | R(%) |
| Base | 910 | - | 55739 | 0 | 25.6 | - | 26 | 0 |
| RS#1 | 489 | 10 | 32446 | **42.8** | 15.3 | 0.7 | 26.85 | -4.88 |
| RS#2 | | 34 | 36816 | 34.05 | | 1.2 | 27.97 | -9.25 |
| RS#3 | | 55 | 40577 | 27.02 | | 1.8 | 28.89 | -11.11 |
| RS#4 | | 68 | 44768 | 19.69 | | 2.0 | 30.23 | -16.27 |
| RSP#1 | | 10 | 33249 | 40.35 | | 0.7 | 16.72 | **34.69** |
| RSP#2 | | 34 | 38422 | 31.07 | | 1.2 | 17.26 | 32.58 |
| RSP#3 | | 55 | 42987 | 22.88 | | 1.8 | 18.21 | 29.97 |
| RSP#4 | | 68 | 47981 | 13.92 | | 2.0 | 18.83 | 27.58 |

**R(%)**: Reduction ratio compared with Base architecture
**SW**: Bus switch

## 5.3 Performance evaluation

We have applied several kernels of Livermore loops benchmark and representative loops in DSP applications to RS and RSP architectures. Table 3 shows the operation set and maximum multiplication number in a cycle of selected kernels. Table 4 shows that RSP Arch#2 supports all of the selected kernels of Livermore loops without stall and gives the best performance. Table 5 shows performance evaluation of specific applications including 2D-FDCT, SAD, MVM, and FFT, and RSP Arch#2 again supports all of the selected kernels. The performance evaluation results of Table 4 and 5 show that the best performance for individual kernels can be obtained with RSP#1 or RSP#2. Therefore we conclude that RSP architecture is more efficient than the base architecture in the aspect of area and delay. Furthermore, the shared resources of RSP architectures are more utilized than RS architectures under same resource sharing condition - in the case of 2D-FDCT in H.263, RSP#2 has no stall but RS#2 has stall cycles of 6.

**Table 3. Kernels in the experiments**

| Kernels | Operation set | Mult No. |
|---|---|---|
| *Hydro | mult, add | 6 |
| *ICCG | mult, sub | 4 |
| *Tri-diagonal | mult, sub | 4 |
| *Inner product | mult, add | 8 |
| *State | mult, add | 7 |
| 2D-FDCT in H.263 enc | mult, shift, add, sub | 16 |
| SAD in H.263 enc | abs, add | 0 |
| Matrix-vector Multiplication | mult, add | 8 |
| Multiplication Loop in FFT | add, sub, mult | 8 |

*****kernel** : Livermore loop benchmark suite
**Mult No** : maximum of multiplications mapped to array in a cycle
mult : multiplication, add : addition, sub : subtraction,
abs : absolute value, shift : bit shift operation

The amount of performance improvement depends on the application. For example, compared to 2D-FDCT which has multiplications, we have achieved much more performance improvement with RSP architectures for SAD which has no multiplication. This is because the clock frequency has been increased by pipelining the multipliers. We cannot have that much speedup for kernels with many multiplications since multiplications take multiple cycles in the RSP architectures.

## 6 Conclusion

Coarse-grained reconfigurable architectures are considered to be appropriate for embedded systems because it can satisfy both flexibility and high performance. Most reconfigurable architectures use regular structures composed of computational primitives for parallel execution of multiple operations and flexible operation scheduling but such regular designs require many hardware resources without regard to the characteristics of the application domain. To overcome the limitation, we suggest a novel reconfigurable architecture template which splits the computational resources into two groups: primitive resources and critical resources. Critical resources can be area-critical and/or delay-critical. Primitive resources compose the base reconfigurable array. Area-critical resources are shared among the basic PEs. The number of shared resources is configurable according to the application domain. Delay-critical resources can be pipelined to curtail the overall critical path so as to increase the system clock frequency. In this way, our architecture template can be used to achieve the goal of domain-specific optimization while keeping the regularity of the reconfigurable array.

In the experiments, the RTL synthesis results show that our resource sharing and pipelining can reduce the area and the critical path delay by up to 42.8% and 34.69% respectively compared to the base architecture and the benchmark evaluation reveals the performance enhancement up to 35.7%. In this paper, we consider only hardware cost and performance but the domain-specific optimization may also be effective for reducing power consumption. To verify the practicality of the proposed approach more seriously, we are currently working on implementing H.264 encoder on our architecture template.

## Acknowledgements

This work was supported by grant No. R01-2004-000-10268-0 from the Basic Research Program of the Korea Science & Engineering Foundation.

### Table 4. Performance evaluation of the kernels in Livermore loop benchmark suite

| Arch' | Hydro(32[†]) | | | | ICCG(32[†]) | | | | Tri-diagonal(64[†]) | | | | Inner product(128[†]) | | | | State(16[†]) | | | |
|---|---|---|---|---|---|---|---|---|---|---|---|---|---|---|---|---|---|---|---|---|
| | cycle | ET(ns) | DR(%) | stall | cycle | ET(ns) | DR(%) | stall | cycle | ET(ns) | DR(%) | stall | cycle | ET(ns) | DR(%) | stall | cycle | ET(ns) | DR(%) | stall |
| **Base** | 15 | 390 | 0 | - | 18 | 468 | 0 | - | 17 | 442 | 0 | - | 21 | 546 | 0 | - | 20 | 520 | 0 | - |
| **RS#1** | 19 | 510.15 | -30.80 | 4 | 18 | 483.3 | -3.26 | 0 | 17 | 456.45 | -3.26 | 0 | 21 | 563.85 | -3.26 | 0 | 35 | 939.75 | -80.72 | 15 |
| **RS#2** | 15 | 419.55 | -1.07 | 0 | 18 | 503.46 | -7.58 | 0 | 17 | 475.49 | -7.58 | 0 | 21 | 587.37 | -7.58 | 0 | 20 | 559.4 | -7.58 | 0 |
| **RS#3** | 15 | 433.35 | -11.11 | 0 | 18 | 520.02 | -11.11 | 0 | 17 | 491.13 | -11.11 | 0 | 21 | 606.69 | -11.11 | 0 | 20 | 577.8 | -11.11 | 0 |
| **RS#4** | 15 | 453.45 | -16.27 | 0 | 18 | 544.14 | -16.27 | 0 | 17 | 513.91 | -16.27 | 0 | 21 | 634.83 | -16.27 | 0 | 20 | 604.6 | -16.27 | 0 |
| **RSP#1** | 21 | 351.12 | 10 | 2 | 19 | 317.68 | 32.12 | 0 | 18 | 300.96 | **31.91** | 0 | 22 | 367.84 | **32.64** | 0 | 37 | 618.64 | -18.96 | 14 |
| **RSP#2** | 19 | 327.94 | **15.92** | 0 | 19 | 327.94 | 29.93 | 0 | 18 | 310.68 | 29.71 | 0 | 22 | 379.72 | 30.45 | 0 | 23 | 396.68 | **23.65** | 0 |
| **RSP#3** | 19 | 345.99 | 11.28 | 0 | 19 | 345.99 | 26.07 | 0 | 18 | 327.78 | 25.84 | 0 | 22 | 400.62 | 26.62 | 0 | 23 | 418.83 | 19.45 | 0 |
| **RSP#4** | 19 | 357.77 | 8.26 | 0 | 19 | 357.77 | 23.55 | 0 | 18 | 338.94 | 23.31 | 0 | 22 | 414.26 | 24.12 | 0 | 23 | 433.09 | 16.71 | 0 |

**Kernel(No.[†])** : iteration number of the kernel, **ET(ns)** : Execution Time = cycle × Critical path delay(ns), **DR(%)** : Delay Reduction percentage,
**stall** : stall number of resource lack, **bold number** : the largest amount of delay reduction %

### Table 5. Performance evaluation of 2D-FDCT, SAD, MVM and FFT function

| Arch' | 2D-FDCT in H.263 enc | | | | SAD in H.263 enc | | | | *MVM(64[†]) | | | | Multiplication Loop in FFT(32[†]) | | | |
|---|---|---|---|---|---|---|---|---|---|---|---|---|---|---|---|---|
| | cycle | ET(ns) | DR(%) | stall | cycle | ET(ns) | DR(%) | stall | cycle | ET(ns) | DR(%) | stall | cycle | ET(ns) | DR(%) | stall |
| **Base** | 32 | 832 | 0 | - | 39 | 1014 | 0 | - | 19 | 494 | 0 | - | 23 | 598 | 0 | - |
| **RS#1** | 56 | 1503.6 | -80.72 | 24 | 39 | 1047.15 | -3.26 | 0 | 19 | 510.15 | -3.26 | 0 | 37 | 993.45 | -66.12 | 14 |
| **RS#2** | 38 | 1062.86 | -7.58 | 6 | 39 | 1090.83 | -7.58 | 0 | 19 | 531.43 | -7.58 | 0 | 23 | 643.31 | -7.58 | 0 |
| **RS#3** | 32 | 924.48 | -11.11 | 0 | 39 | 1126.7 | -11.11 | 0 | 19 | 548.91 | -11.11 | 0 | 23 | 664.47 | -11.11 | 0 |
| **RS#4** | 32 | 967.36 | -16.27 | 0 | 39 | 1178.97 | -16.27 | 0 | 19 | 574.37 | -16.27 | 0 | 23 | 695.29 | -16.27 | 0 |
| **RSP#1** | 64 | 1070.08 | -28.61 | 24 | 39 | 652.08 | **35.7** | 0 | 20 | 334.4 | **32.31** | 0 | 40 | 668.8 | -11.83 | 13 |
| **RSP#2** | 40 | 690.4 | **17.01** | 0 | 39 | 673.14 | 33.61 | 0 | 20 | 345.2 | 30.12 | 0 | 27 | 466.02 | **22.07** | 0 |
| **RSP#3** | 40 | 728.4 | 12.45 | 0 | 39 | 710.19 | 29.96 | 0 | 20 | 364.2 | 26.27 | 0 | 27 | 491.67 | 17.78 | 0 |
| **RSP#4** | 40 | 753.2 | 9.47 | 0 | 39 | 734.37 | 27.57 | 0 | 20 | 376.6 | 23.76 | 0 | 27 | 508.41 | 14.98 | 0 |

*MVM :** Matrix Vector Multiplication